\documentclass[12pt]{article}
\usepackage{amsmath,amssymb}
\usepackage[english]{babel}
\usepackage[cp866]{inputenc}
\usepackage{hyperref}

\numberwithin{equation}{section}

\newcommand{\bq}{\begin{eqnarray}}
\newcommand{\eq}{\end{eqnarray}}
\newcommand{\bbq}{\begin{equation*}}
\newcommand{\eeq}{\end{equation*}}
\newcommand{\ra}{\rightarrow}
\newcommand{\ov}{\overline}

\newcommand{\la}{\Lambda_Q}

\newcommand{\lym}{\Lambda_{SYM}}

\begin{document}

\begin{center} {\bf \large Phase transitions between confinement and higgs phases \\ in ${\cal N}=1\,\, SU(N_c)$ SQCD with $1\leq N_F\leq N_c-1$ quark flavors}.                                      \end{center}
\vspace*{2mm}
\begin{center}\bf Victor L. Chernyak $^{\,a,\,b}$ \end{center}
\begin{center}(e-mail: v.l.chernyak@inp.nsk.su) \end{center}
\begin{center} $^a\,$ Novosibirsk State University,\\ 630090 Novosibirsk, Pirogova str.2, Russia\end{center}
\begin{center} $^b\,$ Budker Institute of Nuclear Physics SB RAS, \\
 630090 Novosibirsk, Lavrent'ev ave.11, Russia\end{center}

\begin{center}{\bf Abstract} \end{center}

\vspace*{2mm}

Considered is 4-dimensional  ${\cal N}=1$ supersymmetric $SU(N_c)$ QCD (SQCD) with $1\leq N_F\leq N_c-1$ equal mass quark flavors in the fundamental representation. The
gauge invariant order parameter $\rho$ is introduced distinguishing confinement (with $\rho=0$) and higgs (with $\rho\neq 0$) phases.

Using a number of independent arguments for different variants of transition between the confinement and higgs regimes, it is shown that transitions between these  regimes are not crossovers but the phase transitions. Besides, it is argued that these phase transitions are of the first order.

This is opposite to the conclusion of the E. Fradkin and S.H. Shenker paper [10] that the transition between the confinement and higgs regimes is the
crossover, not the phase transition. And although the theories considered in this paper and in [10] are different, an experience shows that there is a widely
spread opinion that the conclusion of [10] is applicable to all QCD-like theories: both lattice and continuum, both not supersymmetric and supersymmetric.
This opinion is in contradiction with the results of this paper.

\tableofcontents
\numberwithin{equation}{section}

\section{Introduction}

\hspace*{4mm} Considered is the standard ${\cal N}=1$ SQCD with $SU(N_c)$ colors and $1\leq N_F\leq N_c-1$ flavors of equal mass quarks with the mass parameter $m_Q=m_(\mu=\la)$ in the Lagrangian, where $\la$ is the scale factor of the gauge coupling in the UV region, see e.g. \cite{ADS},\cite{S2}.

The purpose of this paper is to show that, in this theory with light quarks with fixed $m_Q\ll\la$, there is {\it the phase transition} from the region of not
too large $N_c$ where all quarks are higgsed with $\mu_{\rm gl}\gg\la$ \eqref{(2.1)}, with the gauge invariant order parameter $\rho_{\rm higgs}\neq 0$, to
the region of sufficiently large $N_c$ \eqref{(3.1)} where $\rho_{HQ}=0$ and all quarks are in the HQ (heavy quark) phase and are not higgsed but confined.

Besides, we show in section 4.1 that, at fixed $N_c$, there is {\it the phase transition} from the region $m_Q\gg\la$ where all quarks are not higgsed but confined and $\rho_{HQ}=0$, to the region of sufficiently small $m_Q\ll\la$ where they all are higgsed with $\rho_{\rm higgs}\neq 0$.

And finally, using independent arguments based on realization of the flavor symmetry $SU(N_F)$, we show in section 4.2 that, at fixed $N_c,\,\,N_F=N_c-1$, there is {\it the phase transition} from the region $m_Q\gg\la$ where all quarks are not higgsed but confined to the region of sufficiently small $m_Q\ll\la$ where they all are higgsed and not confined.\\

For all this, let us recall first in short some properties of the standard ${\cal N}=1$ SQCD with $SU(N_c)$ colors and $1\leq N_F <3 N_c$ flavors of light
equal mass quarks, see e.g. section 2 in \cite{ch1}. It is convenient to start e.g. with $3 N_c/2 < N_F < 3N_c$ and the scale $\mu=\la$. The Lagrangian looks 
as
\footnote{\,
The gluon exponents are implied in Kahler terms.
}
\bq
K={\rm Tr}\,\Bigl (Q^\dagger Q+ (Q\ra {\ov Q}) \Bigr )\,, \quad {\cal W}=-\frac{2\pi}{\alpha(\mu=\la)} S+
m_Q {\rm Tr}\,({\ov Q} Q)\,.   \label{(1.1)}
\eq
Here\,: $m_Q=m_Q(\mu=\la)$ is the mass parameter (it is taken as real positive), $S=\sum_{A,\beta} W^{A,\,\beta}W^{A}_{\beta}/32\pi^2$, where $W^A_{\beta}$ is the gauge field strength, $A=1...N_c^2-1,\, \beta=1,2$,\, $a(\mu)=N_c g^2(\mu)/8\pi^2=N_c\alpha(\mu)/2\pi$ is the gauge coupling with its scale factor $\la$. Let us take now  $m_Q\ra 0$ and evolve to the UV Pauli-Villars (PV) scale $\mu_{PV}$ to define the parent UV theory. The only change in comparison with \eqref{(1.1)} will be the appearance of the corresponding logarithmic renormalization factor $z(\la, \mu_{PV})\gg 1$ in the Kahler term for {\it massless} quarks and
the logarithmic evolution of the gauge coupling: $\alpha(\mu=\la) \ra\alpha(\mu=\mu_{PV})\ll \alpha(\mu=\la)$, while the scale factor $\la$ of the gauge coupling remains the same. Now, we continue the parameter $m_Q$ from zero to some nonzero value, e.g. $0 < m_Q\ll\la$. And this will be {\it a definition} of our parent UV theory.

The Konishi anomaly \cite{Konishi} for this theory looks as  (everywhere below the repeated indices are summed)
\bq
\quad m_Q(\mu)=z^{-1}_Q(\la, \mu)m_Q\,,\quad m_Q\equiv m_Q(\mu=\la)\,,\quad M^i_j(\mu)= z_Q(\la, \mu) M^i_j\,,\quad M^i_j\equiv M^i_j (\mu=\la)\,.
\label{(1.2)}
\eq
\bbq
m_Q(\mu)\langle M^i_j(\mu)\rangle=\delta^i_j \langle S\rangle\,,\quad i,j=1...N_F\,,\quad \langle M^i_j\rangle=\langle{\ov Q}^\alpha_j Q^i_\alpha\rangle
=\delta^i_j\langle M\rangle\,,\quad \alpha=1...N_c\,.
\eeq
Evolving now to lower energies, the regime is conformal at $m^{\rm pole}_Q < \mu<\la$ and the perturbative pole mass of quarks looks as
~\footnote{
\, Here and below we use the perturbatively exact  NSVZ $\beta$-function \cite{NSVZ}. In \eqref{(1.3)} and below $A\sim B$ means equality up to a constant factor independent of $m_Q$ and $N_c$.
}
\bq
m^{\rm pole}_Q=\frac{m_Q}{z_Q(\la,m^{\rm pole}_Q )}\sim \la\Bigl (\frac{m_Q}{\la}\Bigr )^{\frac{N_F}{3N_c}}
\ll\la\,,\quad z_Q(\la, \mu\ll\la)\sim\Bigl (\frac{\mu}{\la} \Bigr )^{\gamma_Q^
{\rm conf}=\frac{3N_c-N_F}{N_F}}\ll 1\,. \label{(1.3)}
\eq

Integrating then inclusively all quarks as heavy at $\mu < m^{\rm pole}_Q$ \eqref{(1.3)},
\footnote{
\,As well known, the global flavor symmetry $SU(N_F)$ is not broken spontaneously in ${\cal N}=1$\, $SU(N_c)$ SQCD for equal mass quarks, see \eqref{(1.2)}.
Therefore, due to the rank restriction at $N_F > N_c\,$, all quarks are not higgsed but confined.
}
there remains $SU(N_c)$ SYM with the scale factor $\lym$ \eqref{(1.4)} of its coupling. Integrating then all gluons via the Veneziano-Yankielowicz (VY) procedure \cite{VY},
one obtains the gluino condensate, see \eqref{(1.2)}
\bq
\lym=\Bigl (\la^{3N_c-N_F}m^{N_F}_Q \Bigr )_{,}^{\frac{1}{3N_c}}\quad \langle S\rangle=\lym^3=\Bigl (\la^{3N_c-N_F}m^{N_F}_Q \Bigr )^{\frac{1}{N_c}}=m_Q
\langle M \rangle,\,\, \langle M \rangle=\la^2\Bigl (\frac{m_Q}{\la} \Bigr )_{.}^{\frac{N_F-N_c}{N_c}}\,  \label{(1.4)}
\eq

Another way, we can take the IR-free ${\cal N}=1$ SQCD with $N_F > 3N_c$ and to start from $\mu=\la$ with $m_Q\ll\la$. All quarks are not higgsed but confined and decouple as heavy at $\lym\ll\mu=m^{\rm pole}_Q=m_Q/z_Q(\la, m^{\rm pole}_Q)\ll\la$ {\it in the IR-free logarithmic weak coupling regime}, where $z_Q(\la, m^{\rm pole}_Q)\ll 1$ is the logarithmic renormalization factor. There remains ${\cal N}=1\,\, SU(N_c)$ SYM with the scale factor $\lym$ of its coupling. From matching the couplings $a_{+}(m^{\rm pole}_Q)=a_{SYM}(m^{\rm pole}_Q)$ one obtains the same $\lym$ \eqref{(1.4)}.

Finally, let us consider the region $N_c< N_F<3 N_c/2$, where light quarks with $m_Q\ll\la$ are in the (very) strong coupling regime with
$a(\mu\ll\la)\sim (\la/\mu)^{(3N_c-2N_F)/(N_F-N_c)}\gg1$ at $\mu\ll\la$, see section 7 in \cite{ch1}. The quark perturbative pole mass looks in this case as, see
section 7 in \cite{ch1}
\bq
\lym\ll m^{\rm pole}_Q=\frac{m_Q}{z_Q(\la,m^{\rm pole}_Q )}\sim \la\Bigl (\frac{m_Q}{\la} \Bigr )^{\frac{N_F-N_c}{N_c}}\ll\la,\,\, z_Q(\la, \mu\ll\la)\sim \Bigl (\frac{\mu}{\la} \Bigr )^{\gamma_Q^{str}=\frac{2N_c-N_F}{N_F-N_c}}\ll 1\,.\,\,\, \label{(1.5)}
\eq
Integrating inclusively all quarks as heavy at $\mu < m^{\rm pole}_Q$ (see footnote 3), there ramains $SU(N_c)$ SYM with the scale factor $\lym$. From matching the couplings $a_{+}(m^{\rm pole}_Q)=a_{SYM}(m^{\rm pole}_Q)$, see \eqref{(1.5)},\eqref{(3.3)},\eqref{(3.7)}, one obtains the same $\lym$ \eqref{(1.4)}
\bq
a_{+}(\mu=m^{\rm pole}_Q)=\Bigl (\frac{\la}{m^{\rm pole}_Q} \Bigr )^{\nu=\frac{3N_c-2N_F}{N_F-N_c}}=a^{(str,\, pert)}_{SYM}(\mu=m^{\rm pole}_Q)=\Bigl (\frac{m^{\rm pole}_Q)}{\lym} \Bigr )^3\,\ra\, \lym=\Bigl (\la^{3N_c-N_F}m^{N_F}_Q \Bigr )_{,}^{1/3N_c}\, \label{(1.6)}
\eq
as it should be.\\

Now \eqref{(1.4)} can be continued to $1 \leq N_F < N_c$ considered in this paper.\\

\section{The Higgs phase}

In this range $1 \leq N_F < N_c$, the weak coupling Higgs phase at $\mu_{\rm gl}\gg\la$ for light quarks with $0 < m_Q\ll\la$ looks as follows, see e.g. section 2 in \cite{ch1}. All quarks are higgsed, i.e. form a constant coherent condensate in a vacuum state, at the high scale $\mu\sim\mu_{\rm gl}\gg\la$ in the logarithmic weak coupling regime.
~\footnote{
\,We ignore from now on for simplicity all logarithmic factors and trace only the power dependence on $m_Q/\la$ and $N_c$.
}
And the perturbative pole masses of $N_F(2N_c-N_F)$ massive gluons look as
\bq
\Bigl (\frac{\mu_{\rm gl}}{\la}\Bigr )^2\sim g^2(\mu=\mu_{\rm gl})\,z_Q(\la,\mu=\mu_{\rm gl})\, \frac{\rho_{\rm higgs}^2} {\la^2}
\sim\frac{1}{N_c}\frac{\langle M\rangle}{\la^2}\sim\frac{1}{N_c}\frac{\langle S\rangle}{m_Q\la^2}
\sim \frac{1}{N_c}\Bigl (\frac{\la}{m_Q} \Bigr )^{\frac{N_c-N_F}{N_c}}\gg 1\,. \label{(2.1)}
\eq
\bbq
\quad  g^2(\mu=\mu_{\rm gl})\approx \frac{8\pi^2}{(3N_c-N_F)\ln\Bigl (\mu_{\rm gl}/\la\Bigr )}\sim \frac{1}{N_c}\,,\quad
z_Q(\la,\mu=\mu_{\rm gl})\sim\Bigl (\ln\frac{\mu_{\rm gl}}{\la}\Bigr )^{\frac{N_c}{3N_c-N_F}}\sim 1\,.
\eeq

Higgsing of all $N_F$ quarks with $1\leq N_F\leq N_c-2$ flavors at $\rho_{\rm higgs}=\la\Bigl (\frac{\la}{m_Q} \Bigr )^{\frac{N_c-N_F}{2N_c}}\gg\la$ \eqref{(1.4)} breaks spontaneously separately the global $SU(N_F)$ and global  $SU(N_c)\ra SU(N_c-N_f)$, but there remains unbroken diagonal $SU(N_F)_{C+F}$ global symmetry. Besides, the gluons from remained $SU(N_c-N_F)$ SYM do not receive large masses $\sim\rho_{higgs}$ and remain (effectively) massless at scales $\mu>\lym$.\\

Dealing with higgsed quarks, to obtain \eqref{(2.1)}, we first separate out Goldstone fields from quark fields $Q^i_{\alpha}$ normalized at the scale $\la$ (a part of these Goldstone fields or all of them will be eaten by gluons  when quarks are higgsed)
\bq
Q^i_{\alpha}(x)=\Bigl (V_{\rm Goldst}^{SU(N_c)}(x)\Bigr )^{\beta}_{\alpha}\,{\hat Q}^i_{\beta}(x)\,,\quad {\hat Q}^i_{\beta}(x)=\Bigl (
V_{\rm Goldst}^{SU(N_c)}(x)^\dagger \Bigr )^\gamma_\beta Q^i_{\gamma}(x)\,, \label{(2.2)}
\eq
\bbq
{\hat Q}^i_{\beta}(x)=\Bigl (  U_{\rm global}^{SU(N_c)}\Bigr )^{\delta}_{\beta} \Bigl ( U_{\rm global}^{SU(N_F)}\Bigr )^i_j {\tilde Q}^j_{\delta}(x),
\quad \alpha, \beta, \gamma, \delta=1...N_c,\,\, i, j=1...N_F,
\eeq
where $V_{\rm Goldst}^{SU(N_c)}(x)$ is the $N_c\times N_c$ unitary $SU(N_c)$ matrix of Goldstone fields.

But the physical degrees of freedom of massive scalar superpartners of massive gluons and light pion fields $\Pi^i_j$ \eqref{(2.5)}, originating from combined physical degrees of freedom of ${\hat Q}$ and ${\hat{\ov Q}}$, remain in ${\hat Q}^i_{\beta}(x)$ and ${\hat {\ov Q}_i^{\,\beta}(x)}$.

This can be checked by direct counting. The quark fields $Q$ and ${\ov Q}$ have $4 N_F N_c$ real physical degrees of freedom  on the whole. From these, $N_F(2N_c-N_F)$ real Goldstone degrees of freedom are eaten by massive gluons. (The extra $(N_c-N_F)^2-1$ real Goldstone modes not eaten by remaining perturbatively massless $SU(N_c-N_F)$ gluons remain not physical due to the gauge invariance of the Lagrangian). The same number $N_F(2N_c-N_F)$ of combined real physical degrees of freedom of $Q$ and ${\ov Q}$ form scalar superpartners of massive gluons. And remaining $2 N_F^2$ combined real degrees of freedom of $Q$ and ${\ov Q}$ form $N_F^2$ complex physical degrees of freedom of light colorless mesons $M^i_j$.

And then, with the standard choice of vacuum of spontaneously broken global symmetry, we replace ${\hat Q}^i_{\beta}(x)$ in \eqref{(2.2)}, containing remained degrees of freedom, by its mean vacuum value (at $\mu=\la$)
\bq
\langle {\hat Q}^i_{\beta}(x)\rangle =\langle {\hat Q}^i_{\beta}(0)\rangle=\delta^i_{\beta}\,\rho_{\rm higgs}\,, \quad \frac{\rho_{\rm higgs}}{\la}=\Bigl (\frac{\la}{m_Q}\,\Bigr )^{\frac{N_c-N_F}{2 N_c}}\gg 1\,,\quad i=1...N_F\,,\quad \beta=1...N_c\,. \label{(2.3)}
\eq
And similarly $\langle {\hat {\ov Q}_i^{\,\beta}(x)}\rangle=\delta_i^{\,\beta}\,\rho_{\rm higgs}$.

Under pure gauge transformations, see \eqref{(2.2)}\,:
\bq
Q^i_\alpha (x)\ra  \Bigl ( V_{\rm pure\, gauge}^{SU(N_c)}(x)\Bigr )^\beta_\alpha Q^i_\beta (x),\quad \quad
V_{\rm Goldst}^{SU(N_c)}(x)\ra V_{\rm pure\, gauge}^{SU(N_c)}(x) V_{\rm Goldst}^{SU(N_c)}(x). \label{(2.4)}
\eq
That is, these are $Q^i_\alpha (x)$ and Goldstone fields which are transformed in \eqref{(2.2)},\eqref{(2.4)},  while ${\hat Q}^i_{\beta}(x)$ stays intact under pure gauge transformations and is the gauge invariant quark field.   And $\rho_{\rm higgs}\neq 0$ in \eqref{(2.3)} is the {\it gauge invariant order parameter} for higgsed scalar quarks, while $\rho_{\rm HQ}=0$ if quarks are in the HQ (heavy quark) phase and not higgsed, see \eqref{(4.1.1)},\eqref{(4.1.2)}. ( And, in particular, $V_{\rm pure\, gauge}^{SU(N_c)}(x) V_{\rm Goldst}^{SU(N_c)}(x)=I_{SU(N_c)}$  in the so called "unitary gauge"\, where $I_{SU(N_c)}$ is the unity matrix).\\

Under the replacement $Q\ra V_{\rm Goldst}^{SU(N_c)}(x){\hat Q}$ \eqref{(2.2)}, the covariant derivative $i D_{\nu}(A)Q=(i\partial_{\nu}+A_{\nu})Q$ is replaced by $V_{\rm Goldst}^{SU(N_c)}(x) i D_{\nu}(B){\hat Q}$, with $B_{\nu}(x)=\Bigl [ \Bigl ( V_{\rm Goldst}^{SU(N_c)}(x)\Bigr )^{\dagger}  A_{\nu}(x) V_{\rm Goldst}^{SU(N_c)}(x)+i\Bigl ( V_{\rm Goldst}^{SU(N_c)}(x)\Bigr )^{\dagger} \partial_{\nu} V_{\rm Goldst}^{SU(N_c)}(x)\Bigr ]$. Now, the fields ${\hat Q}$ and $B_{\nu}$ are invariant under $SU(N_c)$ pure gauge transformations \eqref{(2.4)}.

When all $N_F$ quarks $Q$ and $\ov Q$ are higgsed, $N_F(2N_c-N_F)$ Goldstone modes in $V_{\rm Goldst}^{SU(N_c)}(x)$ \eqref{(2.2)} are eaten by gluons.
The remaining light $SU(N_c-N_f)$ SYM gluon fields $B_{\nu}$ are gauge invariant with respect to the original $SU(N_c)$ gauge transformations. But there appears the emergent standard $SU(N_c-N_F)$ gauge invariance of the lower energy $SU(N_c-N_F)$ SYM Lagrangian $L_{SYM}(B_{\nu})$ written in terms of fields $B_{\nu}(x)$.

The gauge invariant  pole masses of massive gluons are as in \eqref{(2.1)}.\\

{\it The  gauge invariant order parameter $\rho_{\rm higgs}\neq 0$ in \eqref{(2.3)} is the counter-example to a widely spread opinion that the gauge invariant order parameter for higgsed scalar quarks in the fundamental representation does not exist}. Besides, as pointed out in section (6.2) in \cite{ch17}, the attempt to use as the gauge invariant order parameter the mean vacuum value of the {\it colorless} composite operator $\langle M\rangle^{1/2}$ \eqref{(1.2)},
instead of the gauge invariant but {\it colorful} $\langle {\hat Q}^i_{\beta}\rangle=\delta^i_{\beta}\rho$, is erroneous. The reason is that $\langle M\rangle^{1/2} > 0$  \eqref{(1.2)} is small but nonzero due to quantum loop and nonperturbative effects even for heavy quarks with $m_Q\gg\la,\,\,N_F<N_c$, see e.g. \eqref{(1.4)}. Such quarks are in the HQ(=heavy quark)-phase and {\it they are not higgsed really, i.e. $\rho_{\rm HQ}=0,\,\,\rho_{\rm HQ}\neq \langle M\rangle^{1/2}$, see \eqref{(4.1.1)},\eqref{(4.1.2)}}. Or e.g., using in lattice calculations for QCD-like not supersymmetric theories with {\it heavy scalar  not higgsed quarks} the mean vacuum value of the {\it colorless} gauge invariant composite operator $"V"=\langle\, \sum_{\alpha=1}^{N_c}\sum_{i=1}^{N_F} (\phi^\dagger)^\alpha_i \phi^i_\alpha\,\rangle^{1/2} > 0$  as the order parameter (instead of $\langle {\hat \phi}^{\,i}_\alpha\rangle=0$ for such heavy confined quarks). This is also misleading because $"V"\neq 0$ in all regimes due to various quantum effects.

Using  $\langle M\rangle^{1/2}$  \eqref{(1.2)} or $"V"$ as order parameters creates an illusion that the transition between the confinement and higgs regimes is the analytic crossover, while it is really the non-analytic phase transition.

Unlike the {\it analytical} dependence of mean vacuum values of lowest components of {\it colorless} chiral superfields, e.g. $\sum_{\beta=1}^{N_c}\langle{\ov Q}^\beta_j Q^i_\beta\rangle=\delta^i_j M(N_c, N_F,  m_{Q,i})$, on parameters of the superpotential, the mean vacuum values of lowest components of gauge invariant but {\it colorful} chiral superfields, e.g. $\langle{\hat Q}^i_\beta\rangle=\delta^i_{\beta}\,\rho(N_c, N_F, m_{Q, i})$, depend {\it non-analytically} on these parameters. E.g., $\rho$ is nonzero at fixed $N_c$ and sufficiently light quarks but zero for either sufficiently heavy quarks, or for light quarks and sufficiently large $N_c$.

That the order parameter is $\langle {\hat Q}^i_{\beta}\rangle$ and not  $\langle M\rangle^{1/2}$ is especially clearly seen in D-terms of fermions of the Lagrangian \eqref{(1.1)}\,: $\Bigl\{ (Q^\dagger)_i^\beta\,\lambda_\beta^\gamma\,\chi^i_\gamma\,\, + {\rm h.c.}\Bigr\}+(Q\ra {\ov Q})$. The nonzero mass term  
of fermions (superpartners of massive bosons due to higgsed quarks) expressed in terms  gauge invariant fields with hats, see \eqref{(2.2)},\eqref{(2.3)}, looks then as\,: $\sim\Bigl\{ [\,\langle\, ({\hat Q}^\dagger)_i^\sigma\,\rangle=\delta_i^{\sigma}\,\rho_{\rm higgs}\neq0\,]\,{\hat\lambda}_\sigma^\tau\,
{\hat\chi}^{\,i}_\tau\,\, + {\rm h.c.}\Bigr \}+({\hat Q}\ra {\hat{\ov Q}}),\,\, {\hat\lambda}=\Bigl ( V^{SU(N_c)}_{\rm Goldst}(x)\Bigr )^{\dagger}\lambda \Bigl ( V^{SU(N_c)}_{\rm Goldst}(x)\Bigr )$,  where $\chi$ is the fermionic superpartner of $Q$.\\

At $1\leq N_F\leq N_c-2$, due to higgsed quarks, $N_F(2N_c-N_F)$ gluons and the same number of their ${\cal N}=1$ superpartners acquire masses $\mu_{\rm gl}\gg\la$ and decouple at $\mu<\mu_{\rm gl}$. There remain at lower energies  local ${\cal N}=1$\,\,  $SU(N_c-N_F)$ SYM and $N^2_F$ light complex pion fields $\Pi^i_j(x)\,:\,\,M^i_j(x)=\delta^i_j\langle M\rangle +\Pi^i_j(x),\,\, \langle\Pi^i_j(x)\rangle=0,\,\, i,j=1...N_F$. After integrating out all heavy particles with masses $\sim \mu_{\rm gl}\gg\la$, the scale factor of $SU(N_c-N_F)$ SYM looks as, see section 2 in \cite{ch1} and \eqref{(1.2)}
\bq
\Lambda^3_{SYM}=\Bigl (\frac{\la^{3N_c-N_F}}{\det M} \Bigr )^{\frac{1}{N_c-N_F}}, \quad M^i_j=\langle M^i_j\rangle(\mu=\la)+\Pi^i_j\,.
 \label{(2.5)}
\eq

Lowering energy to $\mu\sim \Lambda_{SYM}$ and integrating all $SU(N_c-N_F)$ gluons via the VY procedure \cite{VY}, the Lagrangian of $N^2_F$ light pions $\Pi^i_j$ looks as
\footnote{
\,The whole $SU(N_c)$ group is higgsed at $N_F=N_c-1$ and all $N_c^2-1$ gluons are heavy. {\it There is no confinement}. The last term in the superpotential \eqref{(2.6)} is then due to the instanton contribution \cite{ADS}. For $1\leq N_F\leq N_c-2$ the instanton contribution to superpotential from the broken part of $SU(N_c)$ is zero due to extra gluino zero modes. The nonperturbative term in the superpotential \eqref{(2.6)} originates from nonperturbative effects in the ${\cal N}=1\,\, SU(N_c-N_F)$ SYM, see section 2 in \cite{ch1} and \cite{VY}.
}
\bq
K_M=2\,z_Q(\la,\mu=\mu_{\rm gl}) {\rm Tr}\, \sqrt{M^\dagger M}\,,\quad {\cal W}_{\Pi}= m_Q{\rm Tr}\,M +(N_c-N_F)\Bigl (\frac{\la^{3N_c-N_F}}{\det M} \Bigr )^{\frac{1}{N_c-N_F}}, \label{(2.6)}
\eq
where $z_Q(\la,\mu=\mu_{\rm gl}\gg\la)\gg 1$ is the quark logarithmic renormalization factor.

From this, $\langle M^i_j\rangle$ and the pion masses are
\bq
\langle M^i_j \rangle=\delta^i_j \la^2\Bigl (\frac{\la}{m_Q} \Bigr )^{\frac{N_c-N_F}{N_c}}\,,\quad
\mu^{\rm pole}(\Pi)=\frac{ m_Q}{z_Q(\la,\mu=\mu_{\rm gl})}\ll\lym\ll\la\,.  \label{(2.7)}
\eq

On the whole. All quarks are higgsed and the mass spectrum at $1\leq N_F\leq N_c-2$  looks as follows.\,  a) $SU(N_F)_{\rm adj}\,\,{\cal N}=1$ multiplet of heavy not confined gluons with the mass \eqref{(2.1)};\, b) one heavy  ${\cal N}=1$ multiplet of $SU(N_F)_{\rm singl}$ not confined gluon with the mass \eqref{(2.1)};\, c) $2N_F(N_c-N_F)$\,  ${\cal N}=1$ multiplets of heavy $SU(N_F)\times SU(N_c-N_F)$ bifundamental gluons (hybrids) with masses \eqref{(2.1)}, which behave as quarks with $N_F$ flavors with respect to confining them not higgsed by quarks ${\cal N}=1$ $SU(N_c-N_F)$ SYM and are weakly coupled and weakly confined, see footnote \ref{(f6)};\, d) a number of  ${\cal N}=1$ $SU(N_c-N_F)$ SYM gluonia with the typical mass scale ${\cal O}(\lym)\ll\la$ \eqref{(1.4)} (except for the case $N_f=N_c-1$);\, e) $N^2_F$ light colorless complex pions $\Pi^i_j$  with masses $\sim m_Q\ll\lym$~ \eqref{(2.7)}.

\section{The heavy quark (HQ) phase}

It is seen from \eqref{(2.1)},\eqref{(2.3)} that  at $\mu\gg\la$ the value of the running gluon mass $\mu_{\rm gl}(\mu)\gg\la$ decreases with increasing $N_c$ and fixed
$(m_Q/\la)\ll 1$.  And at sufficiently large number of colors,
\bq
\frac{N_c}{N_c-N_F}\ln (N_c) \gg \ln(\frac{\la}{m_Q})\gg 1\,, \label{(3.1)}
\eq
$\mu_{\rm gl}(\mu\sim\la)$ will be much smaller than $\la$. This means that even quarks with large $(\rho_{\rm higgs}/\la)=\Bigl (\la/m_Q\,\Bigr )^{(N_c-N_F)/2 N_c}\gg 1$ are not higgsed then in the weak coupling regime at $\mu\gg\la$. And now, at such $N_c$ \eqref{(3.1)}, {\it all quarks and gluons will remain effectively massless in some interval of scales} $\mu_H < \mu < \la$. Recall also that considered ${\cal N}=1$ SQCD is outside the conformal window at $N_F < 3N_c/2$ \cite{S2}. Therefore, to see whether quarks are really able to give by higgsing such a mass to gluons which will stop the perturbative massless RG-evolution, we have to consider the region $\mu\ll\la$ where the theory entered into a {\it perturbative strong coupling regime} with $a(\mu\ll\la)=N_c\alpha(\mu)/2\pi\gg 1$.

Let us recall a similar situation at $N_c < N_F < 3N_c/2$ considered in section 7 of \cite{ch1} (only pages 18 - 21 including the footnote 18 in arXiv:0712.3167\, [hep-th]). As pointed out therein, when decreasing scale $\mu$ crosses $\mu\sim\la$ from above, the increasing perturbative coupling $a(\mu)$ crosses unity from below. But for (effectively) massless quarks and gluons the perturbatively exact NSVZ $\beta$-function \cite{NSVZ}
\bq
\frac{d a(\mu)}{d\ln \mu} = \beta(a)=-\, \frac{a^2}{1-a}\, \frac{(3N_c-N_F)-N_F\gamma_Q(a)} {N_c}\,,\quad a(\mu)=N_c g^2(\mu)/8\pi^2=N_c\alpha(\mu)/2\pi\,  \label{(3.2)}
\eq
{\it can't change its sign by itself (and can't become frozen at zero outside the conformal window) and behaves smoothly}. I.e., when increased $a(\mu)$ crosses unity from below and denominator in \eqref{(3.2)} crosses zero, the increased quark anomalous dimension $\gamma_Q(\mu)$ crosses $(3N_c-N_F)/N_F$ from below, so that the $\beta$-function behaves smoothly and remains negative at $\mu < \la$.  The coupling $a(\mu\ll\la)$ continues
to increase with decreasing $\mu$
\vspace*{-2mm}
\bq
\frac{d a(\mu)}{d\ln \mu} = \beta(a)\ra \, -\, \nu\, a\,<\, 0,\quad  \nu=\Bigl [\frac{N_F}{N_c}(1+\gamma^{\rm str}_Q)-3\Bigr ]={\rm const} > 0\,, \quad                 a(\mu\ll\la)\sim\Bigl (\frac{\la}{\mu} \Bigr )^{\nu\, >\, 0}\gg 1\,. \label{(3.3)}
\eq
In section 7 of \cite{ch1} (see also \cite{ch3},\cite{Session}) the values $\gamma^{\rm str}_Q=(2N_c-N_F)/(N_F-N_c) > 1,\,\,\nu=(3N_c-2N_F)/(N_F-N_c) > 0$ at $\mu\ll\la$ and $N_c < N_F < 3N_c/2$ have been found from matching of definite two point correlators in the direct $SU(N_c)$ theory and in $SU(N_F-N_c)$ Seiberg's dual \cite{S2}. In our case here with $1 \leq N_F < N_c$ the dual theory does not exist. So that, unfortunately, we can't find the concrete value $\gamma^{\rm str}_Q$. But, as will be shown below, for our purposes it will be sufficient to have the only condition $\nu > 0$ in \eqref{(3.3)}.

Let us look now whether, at large $N_c$ \eqref{(3.1)}, a potentially possible higgsing of quarks, even with large $(\rho_{\rm higgs}/\la)=\Bigl (\la/m_Q\,\Bigr )^{(N_c-N_F)/2 N_c}\gg 1$, can give gluons such a mass which will stop the perturbative massless RG-evolution. At large $N_c$ \eqref{(3.1)},  such running gluon mass would look at $\mu\ll\la$ as, see \eqref{(2.1)},\eqref{(3.3)},\eqref{(3.6)},\eqref{(3.7)}
\bq
\frac{\mu^2_{\rm gl}(\mu\ll\la, N_c)}{\mu^2}\sim \frac{a(\mu\ll\la)}{N_c}\, z_Q(\la,\mu\ll\la)\,\frac{\rho_{\rm higgs}^2}{\mu^2}\sim
\Bigl (\frac{\mu}{\la} \ll 1\Bigr )^{\Delta >\, 0 }\Bigl [\frac{1}{N_c}\Bigl (\frac{\la}{m_Q}\Bigr )^{\frac{N_c-N_F}{N_c}}\ll 1 \Bigr ] \ll 1, \,\,\label{(3.4)}
\eq
\bbq
\Delta=\frac{N_c-N_F}{N_c}(1+\gamma^{\rm str}_Q) > 0\,,\quad z_Q(\la,\mu\ll\la)\sim \Bigl (\frac{\mu}{\la}\Bigr )^
{\gamma^{\rm str}_Q\,>\,2}\ll 1\,,\quad {\rm at}\quad  m^{\rm pole}_Q < \mu \ll \la\,,
\eeq
\bq
\hspace*{-1.5mm}\frac{\mu^2_{\rm gl}(\mu < m^{\rm pole}_Q, N_c)}{\mu^2}\sim \frac{a^{(str, pert)}_{SYM}(\mu < m^{\rm pole}_Q)}{N_c} z_Q(\la, m^{\rm pole}_Q)\,\frac{\rho_{\rm higgs}^2}{\mu^2}\sim\frac{1}{N_c} \frac{\mu}{m^{\rm pole}_Q}\ll 1,\,  {\rm at}\,\, \lym < \mu < m^{\rm pole}_Q. \,\,\label{(3.5)}
\eq

It is seen from \eqref{(3.4)},\eqref{(3.5)} that, at fixed $(m_Q/\la)\ll 1$ and large $N_c$  \eqref{(3.1)}, even with large $(\rho_{\rm higgs}/\la)=\Bigl (\la/m_Q\,\Bigr )^{(N_c-N_F)/2 N_c}\gg 1$ \eqref{(2.3)} the potentially possible gluon masses become too small. I.e., with increasing $N_c$, in some vicinity
$\mu\sim\la$ the gluon masses become $\mu_{\rm gl}(\mu\sim\la) < \la$. And this inequality is only strengthened with decreasing $\mu:\,\,\mu_{\rm gl}(\mu\ll\la)\ll \mu$.

And so, {\it the potentially possible gluon mass terms in the Lagrangian from still higgsed quarks become too small and dynamically irrelevant. The gluons become  effectively massless}. I.e.,  potentially higgsed quarks become unable to give such masses to gluons which will stop the perturbative massless RG-evolution (and there is no pole in the gluon propagator).

Moreover, with  fixed $(m_Q/\la)\ll 1$ and increasing $N_c$, the numerical values of $\langle {\hat{\ov Q}}\rangle$ and $\langle {\hat Q}\rangle$
(together with the gluon mass term in the Lagrangian from higgsed quarks) {\it drop then to zero} at some value of increasing $N_c$, somewhere in the region
$\mu_{\rm gl}(\mu\sim\la) \sim \la$. And remain zero at this (or larger) value of $N_c$ at smaller $\mu\,:\,\lym < \mu\ll\la$. {\it The drop of the order parameter $\rho$ from $\rho\neq 0$ to $\rho=0$ is the phase transition}. The physical reason for this is that {\it when gluons become perturbatively
massless (and effectively massless because their nonperturbative masses from SYM are small, $\sim\lym$), the physical, i.e. path dependent, phases of colored quark fields $\hat Q$ and ${\hat{\ov Q}}$  become freely fluctuating due to interactions with such gluons}.

And although  the mean value $\langle M\rangle=\sum_{\alpha=1}^{N_c}\langle{\ov Q}^\alpha_1 Q^1_\alpha\rangle (\mu=\la)=\langle S\rangle/m_Q\gg\la^2$ \eqref{(1.4)} remains the same, {\it it becomes nonfactorizable}  because gluons become (effectively) massless and quarks become unhiggsed. And all this shows that the assumption about higgsed quarks with $\langle{\hat Q}\rangle=\langle{\hat{\ov Q}}\rangle\neq 0$ becomes not self-consistent at fixed $m_Q/\la\ll 1$ and sufficiently large $N_c$ \eqref{(3.1)}.

This regime (i.e. the HQ-phase) with light quarks with fixed $m_Q/\la\ll 1$ and large $N_c$ \eqref{(3.1)} is qualitatively the same as those for heavy quarks with $m_Q/\la\gg 1$, see section 4.1 below. They are also not higgsed, i.e. $\langle{\hat{\ov Q}}\rangle=\langle {\hat Q}\rangle=0$, but confined and decouple as heavy at $\mu < m_Q^{\rm pole}$, in the weak coupling region where gluons are (effectively) massless. And nonzero value of $\langle M\rangle=\sum_{\alpha=1}^{N_c}\langle{\ov Q}^\alpha_1 Q^1_\alpha\rangle (\mu=\la)=\langle S\rangle/m_Q$ \eqref{(1.4)} is also not due to higgsed quarks but arises from the one quark loop Konishi anomaly for quarks in the HQ (heavy quark) phase.

The meaning and properties of the operator $M^i_j$ are very different for higgsed or not higgsed at large $N_c$  \eqref{(3.1)} quarks. While
$M^i_j=[\,\delta^i_j\rho_{\rm higgs}^2=\delta^i_j \la^2 (\la/m_Q)^{(N_c-N_F)/N_c}]+\Pi^i_j,\,\, \langle \Pi^i_j\rangle=0$, where $\Pi^i_j$ is the one-particle operator of the light pion for higgsed quarks, for not higgsed quarks with $\rho_{HQ}=0\,\, M^i_j$ is the two-particle quark operator, its mean value $\langle M^i_j\rangle$  becomes {\it nonfactorizable} and originates from the one quark loop Konishi anomaly, see \eqref{(1.2)},\eqref{(1.4)} and section 4.1.

Therefore, let us look in this case of not higgsed quarks on the increasing with decreasing $\mu < \la$ running quark mass $m_Q(\mu < \la)$ and on possible value of the quark perturbative pole mass. It looks as,  see \eqref{(3.3)}
\bbq
m_Q(\mu\ll\la)=\frac{m_Q}{z_Q(\la,\mu\ll\la)}\,,\quad z_Q(\la,\mu\ll\la)\sim \Bigl (\frac{\mu}{\la} \Bigr )^{\gamma^{\rm str}_Q\, >\, 2}\ll 1\,,
\eeq
\bq
m^{\rm pole}_Q=\frac{m_Q}{z_Q(\la,m^{\rm pole}_Q)}\quad\ra \quad
m^{\rm pole}_Q\sim\la \Bigl (\frac{m_Q}{\la} \Bigr )^{0 \, < \frac{1}{1+\gamma^{\rm str}_Q}\, < \,\frac{1}{3}\,} \ll \la\,. \label{(3.6)}
\eq
As a result, {\it all quarks are not higgsed and decouple as heavy} at $\mu < m^{\rm pole}_Q$. There remains at lower energies the ${\cal N}=1\,\, SU(N_c)$ SYM {\it in the perturbative strong coupling branch}. From the NSVZ $\beta$-function \cite{NSVZ}
\bbq
\frac{d a^{(str,\, pert)}_{SYM}(\mu\gg\lym)}{d\ln\mu}= - \frac{3\,\Bigl (a^{(str,\, pert)}_{SYM}(\mu\gg\lym)\Bigr ) ^2}{1-a^{(str,\, pert)}_{SYM}(\mu\gg\lym)}\ra 3\, a^{(str,\, pert)}_{SYM}(\mu)\,,
\eeq
\bq
a^{(str,\, pert)}_{SYM}(\mu\gg\lym)\sim\Bigl (\frac{\mu}{\lym}\Bigr )^3 \gg 1\,, \quad a^{(str,\, pert)}_{SYM}(\mu\sim\lym)={\cal O}(1)\,. \label{(3.7)}
\eq

The scale factor of $\lym$ of the gauge coupling is determined from matching, see \eqref{(3.3)},\eqref{(3.6)},\eqref{(3.7)}
\bq
a_{+}(\mu=m^{\rm pole}_Q)=\Bigl (\frac{\la}{m^{\rm pole}_Q} \Bigr )^{\nu}=a^{(str,\, pert)}_{SYM}(\mu=m^{\rm pole}_Q)=\Bigl (\frac{m^{\rm pole}_Q)}{\lym} \Bigr )^3\,
\ra\, \lym=\Bigl (\la^{3N_c-N_F}m^{N_F}_Q \Bigr )^{1/3N_c}\,,  \label{(3.8)}
\eq
as it should be, see \eqref{(1.4)}. Besides,  as a check of self-consistency, see \eqref{(3.3)},\eqref{(3.6)},\eqref{(3.8)}
\bq
\Bigl (\frac{\lym}{m^{\rm pole}_Q}\Bigr )^3\sim\Bigl (\frac{m_Q}{\la} \Bigr )^{\omega\, >\, 0}\ll 1\,,\quad \omega=
\frac{\nu > 0}{(1+\gamma^{\rm str}_Q)} > 0 \,.  \label{(3.9)}
\eq
as it should be.  At $\mu < \lym$ the perturbative RG evolution stops due to non-perturbative effects $\sim\lym$ in the pure ${\cal N}=1\,\,\,SU(N_c)$ SYM.

On the whole, the mass spectrum at $1\leq N_F\leq N_c-1$ and large $N_c$ \eqref{(3.1)} looks as follows.\, a)\, All quarks are not higgsed (i.e. $\rho_{HQ}=0$ and their color charges are not screened due to $\rho_{HQ}\neq 0$) but decouple as heavy at $\mu<m^{\rm pole}_Q$ and are weakly confined. There is a number of quarkonia with the typical mass scale ${\cal O} (m^{\rm pole}_Q) \ll\la$ \eqref{(3.6)}\eqref{(3.9)}, with different spins and other quantum numbers. {\it This mass scale is checked by eqs.\eqref{(3.8)},\eqref{(3.9)}. Integrating inclusively all these hadrons (i.e. equivalently, integrating inclusively all quarks as heavy at $\mu= m^{\rm pole}_Q$), we obtain the well known beforehand right value of $\lym$} \eqref{(1.4)}.

The confinement originates from the ${\cal N}=1\,\, SU(N_c)$ SYM and so the typical string tension is $\sigma^{1/2}_{SYM}\sim\lym\ll m^{\rm pole}_Q\ll\la$.
\footnote{
\, There is no confinement in Yukawa-like theories without gauge interactions. Confinement originates {\it only} from (S)YM sector. The ${\cal N}=1$ SYM is the theory with only one dimensional parameter $\lym$. Therefore, it can't give a string tension $\sigma^{1/2}\sim\la$ but only  $\sigma^{1/2}\sim\lym\ll\la$.
\label{(f6)}
}
\,\, \,b)\, There is a number of $SU(N_c)$ gluonia with the typical mass scale $\sim \lym$ \eqref{(3.8)}. It is seen from the above that the mass spectra at $\mu^{\rm pert}_{\rm gl}\gg\la$ \eqref{(2.1)} or $\mu^{non-pert}_{\rm gl}\sim\lym\ll\la$ at large $N_c$  \eqref{(3.1)} are qualitatively different.\\

Now, about a qualitative difference between the analytic crossover and not analytic phase transition. The gauge invariant order parameter for quark higgsing is $\rho_{\rm higgs}\neq 0$ \eqref{(2.3)}. As pointed out below \eqref{(3.5)}, the perturbative mass terms of gluons in the Lagrangian originating from higgsed quarks drop to zero because $\rho_{HQ}=0$ drops to zero at large $N_c$ \eqref{(3.1)}. This is due to freely fluctuating physical quark fields phases from interaction with effectively massless gluons (with small nonperturbative masses $\sim\lym$), see also section 4.1. I.e., {\it quarks become unhiggsed at such} $N_c$. While $\rho_{\rm higgs}\neq 0$ and large at $N_c$ from \eqref{(2.1)}  because the corresponding gluons are heavy, $\mu_{\rm gl}\gg\la$. Therefore, with fixed  $(m_Q/\la)\ll 1$ and increasing $N_c$, there  is the phase transition somewhere in the region $\mu_{\rm gl}(\mu\sim\la)\sim\la$.\\

The additional arguments for a phase  transition between these two regions of $N_c$ (as opposite to an analytical crossover) look as follows.

Let us suppose now that, for the analytical crossover instead of the phase transition, the quarks would remain higgsed at large $N_c$ \eqref{(3.1)}, and even with still sufficiently large  ${\tilde \rho}^{\,2}_{\rm higgs}(\mu\sim\la)\sim\la^2 (\la/m_Q)^{(N_c-N_F)/N_c}$  (i.e. ignoring all given above arguments for $\rho=0$). As can be seen from \eqref{(3.4)},\eqref{(3.5)},\eqref{(3.7)}, ($\mu_{0}^{2}=z_{Q}(\la, m^{\rm pole}_Q)\rho_{higgs}^2\ll\lym^2$, and even $N_c\mu_{\rm gl}^{2}(\mu\sim\lym)/\lym^2\sim \lym/m_Q^{\rm pole}\ll 1$), the {\it additional} effects from supposedly still higgsed quarks will be then parametrically small and dynamically irrelevant for the RG-evolution from $\mu=\la/(\rm several)$ down to $\mu\sim\lym$. So, the RG-evolution in \eqref{(3.4)},\eqref{(3.6)}  will remain valid in the range $m^{\rm pole}_Q < \mu<\la$ where all quarks and gluons are (effectively) massless. And the RG-evolution in \eqref{(3.5)},\eqref{(3.7)} (after quarks decoupled as heavy) will also remain valid in the range $\lym<\mu < m^{\rm pole}_Q$ where all gluons remain (effectively) massless. At $\mu\sim\lym$ the larger nonperturbative effects $\sim\lym$ from ${\cal N}=1\, \, SU(N_c)$ SYM come into a game and stop the perturbative RG-evolution with (effectively) massless gluons.

For the case of decoupled as heavy not higgsed quarks with $\rho_{HQ}=0$ (as described above), a widely spread opinion (supported by lattice calculations) is that the confinement of massive quarks {\it originates from higgsing (i.e. condensation) of magnetically charged solitons in $SU(N_c)$ (S)YM}.

But then, for sufficiently heavy \eqref{(3.9)} but still 'slightly higgsed' quarks giving the supposed electric mass $\mu_{\rm gl}^{2}(\mu\sim\lym)/\lym^2\sim \lym/(N_c m_Q^{\rm pole})\ll 1$ to gluons, the regime would be self-contradictory. There would be then in the whole ${\cal N}=1\, \,SU(N_c)$ SYM {\it simultaneously} such 'slightly higgsed' quarks and higgsed magnetically charged solitons with the much larger condensate $\rho_{\rm magn}\sim\lym$. But these magnetically charged solitons and quarks are mutually nonlocal.  For this reason, such solitons will keep quarks confined and will prevent them from  condensing in the vacuum state.\\

In \cite{FS} the special (not supersymmetric) QCD-like lattice $SU(N_c)$ gauge theory with $N_F=N_c$ flavors of scalar quarks $\Phi^i_{\beta}$ in the fundamental representation was considered. In the unitary gauge, all $N_c^2+1$ remained degrees of freedom of these quarks were replaced by one constant parameter $|v|\,:\,  \Phi^i_{\beta}\, \ra\, \delta ^i_{\beta}|v|,\, \beta=1...N_c,\,\, i=1...N_F=N_c$.\, I.e., all quarks were higgsed {\it by hands} even at small $|v|\neq 0$ and all $N^2_c-1$ gluons received electric masses $g |v|$. The matter potential is zero. All other $N^2_c+1$ quark physical dynamical degrees of freedom were deleted by hands. The region with the large values of $|v|$ was considered as the higgs regime, while those with small $|v|$ as the confinement one. The conclusion of \cite{FS} was that the transition between the higgs and confinement regimes is the analytic crossover, not the non-analytic phase transition.

Let us note that {\it this {\it not supersymmetric} model \cite{FS} with such {\it permanently higgsed by hands even at small $\,\,0 < g|v|\ll\Lambda_{YM}
\sim \Lambda_{QCD}$ non-dynamical scalar quarks $\Phi^i_{\beta}$ looks unphysical and is incompatible  with the normal dynamical} electrically charged scalar quarks with all their physical degrees of freedom}.
\footnote{\,
Unlike ${\cal N}=1$ SQCD, in non-supersymmetric $N_F=N_c$ QCD with normal dynamical scalar quarks with all their degrees of freedom, these scalar quarks are not massless even in the limit $m_Q\ra 0$. They acquire non-perturbative dynamical masses $\sim\Lambda_{YM}\sim\Lambda_{QCD}$. And, connected with this, {\it there is confinement with the string tension} $\sigma^{1/2}\sim\Lambda_{YM}\sim\Lambda_{QCD}$ in this limit.
}
 
As pointed out above, higgsed magnetically charged solitons ensuring confinement in YM, are mutually nonlocal with electrically charged quarks. For this reason, such solitons  with the much larger vacuum condensate  $\rho_{\rm magn}\sim\Lambda_{YM}\sim\Lambda_{QCD}\gg g|v|$ {\it will keep normal dynamical scalar quarks confined and will prevent them from condensing with $|v|\neq 0$ in the vacuum state}.\\

On the whole, we presented a number of arguments that, for quarks with $ 1\leq N_F\leq N_c-1$ and fixed $0 < m_Q/\la\ll 1$, there is not the analytical crossover but the non-analytical phase transition at sufficiently large $N_c$ \eqref{(3.1)} between the phases with higgsed or not higgsed but confined quarks. As was argued above, the perturbative mass term of gluons in the Lagrangian, proportional to the order parameter $\rho_{\rm higgs}$ of higgsed quarks  \eqref{(2.3)}, is nonzero and large at values of $N_c$ in \eqref{(2.1)} and  becomes not simply small but zero at sufficiently large $N_c$ \eqref{(3.1)}, because the order parameter $\rho$ drops to zero and quarks become unhiggsed (see also \eqref{(4.1.1)},\eqref{(4.1.2)}). The nonzero gluon masses $\sim\lym$ are of nonperturbative origin from SYM, not due to higgsed quarks.\\

\section{Additional independent arguments for the phase transitions}

\subsection{The phase transition vs crossover with increasing $N_c$ at fixed $m_Q\ll\la$}.
\numberwithin{equation}{subsection}

As emphasized in the text below \eqref{(3.5)}, for all $1\leq N_F\leq N_c-1$, the HQ(=heavy quark)-phase of not higgsed but confined quarks with $m_Q\ll\la$, large $N_c$ \eqref{(3.1)} and $m_Q^{\rm pole}\gg\lym$ \eqref{(3.6)},\eqref{(3.9)}, is qualitatively the same as the HQ-phase of heavy not higgsed but confined quarks with $m_Q\gg\la$.

For heavy quarks with $m^{\rm pole}_Q\gg\la$ and for light quarks with $m_Q\ll\la$ and large $N_c$ \eqref{(3.1)} in this HQ-phase, we can add also the following.  Let us write, see \eqref{(1.2)},\eqref{(3.6)},\eqref{(3.9)}, the normalization point is $\mu=\mu_{p}=m_Q^{\rm pole}\gg\lym$ :
\bbq
\Bigl [\, Q^i_\alpha(x)\Bigr ]_{\mu_{p}}=\Bigl ( V_{\rm Goldst}(x)\Bigr )^{\beta}_{\alpha}\Bigl [\,\langle{\hat Q}^i_{\beta}(x)\rangle +\delta {\hat Q}^i_\beta(x)\Bigr ]_{\mu_{p}}\,,\quad  \Bigl [\,{\ov Q}_i^{\alpha}(x)\Bigr ]_{\mu_{p}}=\Bigl [\,\langle{\hat{\ov Q}}_i^{\beta}(x)\rangle +\delta {\hat{\ov Q}}_i^\beta(x)\Bigr ]_{\mu_{p}}\Bigl ( V_{\rm Goldst}^{\dagger}(x)\Bigr )_{\beta}^{\alpha}\,,
\eeq
\bq
\langle \delta {\hat Q}^i_\alpha(x)\rangle_{\mu_{p}}=\langle \delta {\hat{\ov Q}}_i^\alpha(x)\rangle_{\mu_{p}}\equiv 0\,. \label{(4.1.1)}
\eq
\bbq
\langle M^i_j(x)\rangle_{\mu_{p}}=\sum_{\alpha=1}^{N_c}\langle {\ov Q}^\alpha_j(x) Q^i_\alpha(x)\rangle_{\mu_{p}}=\Bigl [\,\sum_{\alpha=1}^{N_c}\langle {\hat {\ov Q}}_j^\alpha\rangle_{\mu_{p}} \langle {\hat Q}^i_\alpha\rangle_{\mu_{p}}=\delta^i_j\, \rho_{\rm HQ}^2\,\Bigr ]
 +\sum_{\alpha=1}^{N_c}\langle \delta {\hat{\ov Q}}_j^\alpha(x) \delta {\hat Q}^i_\alpha(x)\rangle_{\mu_{p}}\,,
\eeq
\bbq
\langle {\hat Q}^i_\alpha\rangle_{\mu_{p}}=\delta^i_\alpha\rho_{\rm HQ}\,,\quad \langle {\hat {\ov Q}}^\alpha_j\rangle_{\mu_{p}}=\delta^\alpha_j\rho_{\rm HQ}.
\eeq

Now, the one quark loop contribution of such quarks with $m_Q^{\rm pole}\gg\lym$ \eqref{(3.9)} looks as
\bq
\sum_{\alpha=1}^{N_c}\langle \delta {\hat{\ov Q}}_j^\alpha(x) \delta {\hat Q}^i_\alpha(x)\rangle_{\mu_{p}}=\delta^i_j\,\frac{\langle S\rangle}{m^{\rm pole}_Q}
\quad \ra \quad  \rho_{\rm HQ}=0\,,  \label{(4.1.2)}
\eq
i.e. {\it it saturates the Konishi anomaly} \eqref{(1.2)}. Therefore, the equations \eqref{(4.1.1)},\eqref{(4.1.2)} show that not only quarks with $m_Q\gg\la$, but {\it in the whole region of the HQ-phase \eqref{(3.1)} even light quarks with $m_Q\ll\la$ are not higgsed}:
$\langle {\hat Q}^i_{\alpha}\rangle_{\mu_{p}} =\delta^i_\alpha \rho_{\rm HQ}=0,\,\,\langle {\hat {\ov Q}}_j^{\alpha}\rangle_{\mu_{p}}=\delta_i^{\alpha}
\rho_{\rm HQ}=0$.  This is independent confirmation of presented in section 3 arguments that {\it quarks in the HQ-phase are not higgsed}, i.e. $\rho_{\rm HQ}=0$.
Because, at fixed $N_c$ and sufficiently small $m_Q$, in the whole region  $m_Q\ll\la,\,\,\mu_{\rm gl}\gg\la$ the value of $\rho_{\rm higgs}$ is nonzero and large, see \eqref{(2.1)},\eqref{(2.3)} and \eqref{(4.2.1)}, this shows that {\it the order parameter $\rho$ behaves non-analytically: it is  nonzero in the
region of the higgs phase and zero either at large $m_Q\gg\la$ or at small $m_Q\ll\la$ but large $N_c$ \eqref{(3.1)}  in the region of the HQ-phase}. This independently confirms that there is not the analytic crossover but non-analytic phase transition between regimes of confined or higgsed quarks.

\subsection{The phase transition vs crossover with decreasing $m_Q$ \\ from $m_Q\gg\la$ to $m_Q\ll\la$  at fixed $N_c\,,\,N_F=N_c-1$}.
\numberwithin{equation}{subsection}

\hspace*{-4mm} Heavy quarks with $m_Q\gg\la$ have large masses and small mean vacuum value $\langle M\rangle=\la^2(\la/m_Q)^{\frac{(N_c-N_F=1)}{N_c}}\\ \ll\la^2\ll\lym^2\ll m^2_Q$, see \eqref{(1.2)},\eqref{(1.4)}.  For this reason, they are in the HQ-regime and are not higgsed, i.e. $\langle{\hat{\ov Q}}^{\,\alpha}_i\rangle=\langle{\hat Q}^i_{\alpha}\rangle=0$, see also \eqref{(4.1.1)}, \eqref{(4.1.2)}. {\it They are weakly confined, see footnote \ref{(f6)}},
and decouple as heavy in the weak coupling regime at $\mu < m^{\rm pole}_Q=m_Q/z_Q(\la, m^{\rm pole}_Q)\gg\la$, where $z_Q(\la, m^{\rm pole}_Q)\gg 1$ is the logarithmic renormalization factor. The scale factor $\lym$ \eqref{(1.4)} of remained ${\cal N}=1\,\, SU(N_c)$ SYM is determined from matching of logarithmically small couplings $a_{+}(\mu=m^{\rm pole}_Q)=a_{SYM}(\mu=m^{\rm pole}_Q)$.
\footnote{
The nonzero gluon masses originate only in ${\cal N}=1$ SYM due to nonperturbative effects, their typical scale is $\sim\lym$.
}
The small nonzero value of $\langle M\rangle$ originates from the one quark loop Konishi anomaly \eqref{(1.4)} for quarks in the HQ-phase, not due to "slightly
higgsed"\, heavy quarks.\\

{\it The global $SU(N_F)$ is unbroken}. There is in the spectrum a number of heavy flavored quarkonia with typical masses ${\cal O}(m_Q)\gg\lym$ and different quantum numbers. For instance, the quark-antiquark bound states with different spins and other quantum numbers are in the adjoint or singlet representations of unbroken global $SU(N_F)$. It is important that, due to a confinement, {\it there are no particles in the spectrum in the $SU(N_F)$} (anti)fundamental representation of dimensionality $N_F$. Besides, there are in the spectrum a number of $SU(N_F)$ singlet gluonia  with typical masses $\sim\lym=\la (m_Q/\la)^{N_F/3N_c},\, \la\ll\lym\ll m_Q$. \\

The light quarks with $N_F=N_c-1$ flavors with their $4N_F N_c$ real degrees of freedom have small current masses  $m_Q\ll\la$ and large mean vacuum values $\langle M^i_j\rangle=\sum_{\alpha=1}^{N_c}\langle{\ov Q}^{\,\alpha}_j  Q^i_\alpha\rangle=\delta^i_j\la^2 (\la/m_Q)^{\frac{1}{N_c}}\gg \la^2\gg\lym^2,\,\,\mu_{\rm gl}\gg\la$, see \eqref{(2.1)}. They all are higgsed in this case in the weak coupling region and the whole global color group $SU(N_c)$ is broken. The quark mean vacuum values of $\langle{\hat Q}^i_{\alpha}\rangle$ and  $\langle{\hat{\ov Q}}^{\,{\alpha}}_i\rangle$ look as, see \eqref{(2.3)}\,:
\bq
\langle{\hat Q}^i_\alpha\rangle=\delta^i_\alpha\,\omega,\quad \langle{\hat{\ov Q}}^{\,\alpha}_i\rangle=\delta^\alpha_j\,\omega,\quad \omega=\la(\la/m_Q)^{1/2N_c}\gg \la,\quad i=1...N_F,\quad \alpha=1...N_c\,. \hspace*{2.5cm} \label{(4.2.1)}
\eq

Let us present now the additional independent arguments that, at fixed $N_c,\, N_F=N_c-1$, it is not the crossover but phase transition between regions of $m_Q\gg\la$ and $m_Q\ll\la,\, \mu_{\rm gl} \gg \la$.

From \eqref{(4.2.1)}, the unbroken global symmetry looks now as:\, $SU(N_F)\times SU(N_c)\times U(1)_B\ra SU(N_F)_{F+C}\times U(1)_{\tilde B}$\,, i.e. the color-flavor locking. {\it There is no confinement.} All $N_c^2-1=N_F^2+2 N_F$ heavy gluons (which "ate" $N_F^2+2 N_F$ massless Goldstone degrees of freedom from quarks) and the same number of their ${\cal N}=1$ scalar superpartners acquired large masses \eqref{(2.1)}. They form 2 adjoint representations of $SU(N_F)$ plus two $SU(N_F)$ singlets. Plus, and this is most important, else $2 N_F$ heavy gluons $(A_{\mu})^i_{\alpha=N_c},\, (A_{\mu})^{\alpha=N_c}_{i},\, i=1...N_F$ and $2 N_F$ their ${\cal N}=1$ scalar superpartners. These $4 N_f$ form {\it two fundamental and two antifundamental representations of $SU(N_F)$ with dimensionality $N_F$} each. And finally, there are $N^2_F$ light complex pions $\Pi^i_j,\, i,j=1...N_F$ with small masses $\sim m_Q$ \eqref{(2.7)} which form the adjoint and singlet representations of $SU(N_F)$. {\it Therefore, there are only fixed numbers of particles with fixed quantum numbers in the spectrum}.

The mass terms of $N_c^2-1$ heavy massive gluons in the Lagrangian look as:
\bbq
M^2_{\rm gl} =2  g^2(\mu_{\rm gl})\,z_Q(\la,\mu=\mu_{\rm gl})\,\sum_{i=1}^{N_F}\sum_{\alpha,\gamma=1}^{N_c}\langle{\,\Bigl ({\hat Q}^{\,\dagger}\Bigr )}^{\,\alpha}_i\rangle\Biggl\{\sum_{\beta=1}^{N_c}(A_{\mu})_\alpha^\beta\,(A_{\mu})^\gamma_\beta\Biggr\}\langle\,{\hat Q}^{\,i}_\gamma\,\rangle=
\eeq
\bq
K\,\sum_{i=1}^{N_F}\sum_{\beta=1}^{N_c}(A_{\mu})^i_\beta\,(A_{\mu})_i^\beta\,,
\quad K= 2 g^2(\mu_{\rm gl})\,z_Q(\la,\mu=\mu_{\rm gl})\Biggl ( \omega^2=\la^2(\la/m_Q)^{1/N_c}\Biggr )\gg\la^2\,. \label{(4.2.2)}
\eq

From \eqref{(4.2.2)}, the masses of gluons in different representations of unbroken global $SU(N_F)$ are {\it different}\,:
\bq
\mu^2_{\rm gl}\bigl(SU(N_F)_{\rm adj}\bigr )=K,\,\, \mu^2_{\rm gl}\bigl(SU(N_F)_{\rm singl}\bigr )=\frac{1}{ N_c} K,\,\,
\mu^2_{\rm gl}\bigl(SU(N_F)_{\rm fund}\bigr )=\mu^2_{\rm gl}\bigl(SU(N_F)_{\rm anti-fund}\bigr )=\frac{1}{2} K,\, \label{(4.2.3)}
\eq
(and the same for their scalar superpartners). It is seen from \eqref{(4.2.3)} that $N_c^2-1$ heavy gluons have different masses and do not form one adjoint representation of global $SU(N_c)$ at $N_c > 2$.\\

From comparison of mass spectra properties in regions $m_Q\gg\la$ and $m_Q\ll\la, \mu_{\rm gl} \gg \la$ \eqref{(2.1)} it is seen that, although the unbroken global symmetry $SU(N_F)$ is the same, but {\it realized are its different representations}. In the case of heavy confined quarks in the HQ-phase there are no particles in the spectrum in the (anti)fundamental representation of $SU(N_F)$, while in the case of light higgsed quarks such representations are present. E.g., for fixed $N_c$, we can start with the case of heavy quarks with $m_Q\gg\la$ and to diminish continuously $m_Q$ until $m_Q\ll\la$. And when reaching the appropriately small value of $m_Q$, such that $\mu_{\rm gl} \gtrsim \la$ \eqref{(4.2.2)},\eqref{(4.2.3)}, all quarks become higgsed and the behavior of the mass spectrum under unbroken global $SU(N_F)$ transformations changes discontinuously (because the dimensions of representations can not change continuously). {\it This jump is impossible in the case of crossover (which is analytic), this means the phase transition between the confinement and higgs phases}.\\

In other words. The fraction $R_{\rm fund}$ of particles in the (anti)fundamental representation in the mass spectrum can serve in the case considered as the order parameter. This fraction is zero in the confinement region where quarks with $\rho_{HQ}=0$ \eqref{(4.1.2)} are not higgsed. While this fraction is the {\it nonzero constant}  in the region with higgsed quarks with $\rho_{higgs}=\omega \gg\la$ \eqref{(4.2.1)}. I.e.,  $R_{\rm fund}$ behaves non-analytically.  This is a clear sign of the phase transition, because this fraction would behave analytically in the case of crossover.\\

At the same time, the dependence of bilinear mean vacuum value $\langle M^i_j\rangle=\delta^i_j \langle M\rangle$ \eqref{(1.2)} ,\eqref{(1.4)} on $m_Q$ is analytic, but this does not mean that there can not be the phase transition. The qualitative difference is that $\langle M^i_j\rangle=\sum_{\alpha=1}^{N_c}\langle {\hat {\ov Q}^\alpha_j}\rangle \langle {\hat Q}^i_\alpha\rangle=\delta^i_j\,\omega^2\neq 0$, see \eqref{(4.2.1)}, \eqref{(2.3)}, {\it i.e. factorizes} for higgsed quarks with $m_Q\ll\la,\,\mu_{\rm gl} > \la$ (the order parameter is $\omega=\la (\la/m_Q)^{\frac{1}{2 N_c}}\gg\la$ \eqref{(4.2.1)}\,). While for non-higgsed but weakly confined (see footnote \ref{(f6)}) quarks in the HQ-phase with $m_Q\gg\la$ or with $\lym\ll m^{\rm pole}_Q\ll\la$ and large $N_c$ \eqref{(3.1)},\eqref{(3.9)},  this bilinear mean value $\langle M^i_j\rangle$ {\it becomes non-factorizable. It originates then from the one quark loop Konishi anomaly \eqref{(1.2)},\eqref{(1.4)}}), and all $\langle {\hat{\ov Q}}^a_i\rangle=\langle {\hat Q}^i_a\rangle=0$,
see the text under \eqref{(3.5)} and \eqref{(4.1.1)},\eqref{(4.1.2)}.\\

Let us present now the additional arguments that the above described phase transitions are of the first order. Suppose that, vice versa, they are e.g. of the second order. Then, at fixed $0 < m_Q/\la\ll 1$ and increasing $N_c$, there is a finite width region of $N_c$ around $\mu_{\rm gl}(\mu\sim\la)\sim\la$
where the order parameter $\rho_0(N_c,\,m_Q/\la)=\rho(N_c,\,m_Q/\la)/\la$ changes continuously with $N_c$, e.g. from its large  value $\approx\rho_{\rm higgs}$ in \eqref{(2.3)} in the higgs phase down to zero in the HQ-phase.
\footnote{
The order parameter $\rho_0$ is defined in \eqref{(2.3)},\eqref{(1.2)} at the scale $\mu=\la$ and {\it is independent of the scale factor} $\mu$. The whole dependence of
$\mu_{\rm gl}(\mu)$ on the scale  $\mu$ at fixed $\rho_0=\rho(N_c,\,m_Q/\la)/\la$ originates from the running quark renormalization factor $z_{Q}(\mu)$ and from the running coupling $a(\mu)$, see \eqref{(2.1)},\eqref{(3.3)}-\eqref{(3.7)}.
}
Consider now some vicinity of the point where $\rho_0(N_c,\,m_Q/\la)$ reaches zero. This vicinity is such that $0 < \rho_0(N_c,\,m_Q/\la) < 1 \ll \rho_{\rm higgs}$, with $\rho_{\rm higgs}\gg 1$ from \eqref{(2.3)}. I.e. $\rho_0(N_c,\,m_Q/\la)$ is not large but nonzero within it. And, at fixed value of $N_c$ within this interval, the gluon mass terms in the Lagrangian from higgsed quarks are nonzero and became such that $0 < \mu_{\rm gl}(\mu\sim\la)\ll\la$ (and remain $\mu_{\rm gl}(\mu)\ll\mu$ at smaller $\mu$) and are {\it dynamically irrelevant}.

Then quarks are already confined but nevertheless remain  "slightly higgsed"\, at fixed $0 <\rho_0(N_c,\,m_Q/\la) < 1$. Then, because the gluon mass terms in the Lagrangian from higgsed quarks are too small and dynamically irrelevant, the perturbative RG-evolution to smaller scales down to $\mu\sim\lym$ is still described by \eqref{(3.3)}-\eqref{(3.7)}, the only difference is that fixed $\rho_0(N_c,\,m_Q/\la)$ instead of its large value \eqref{(2.3)} is now much smaller,
$0 < \rho_0(N_c,\,m_Q/\la) < 1$.

As was argued in section 3, this variant with confined and simultaneously  "slightly higgsed"\, quarks with fixed $\rho_0(N_c,\,m_Q/\la)\neq
0$ is self-contradictory for the vacuum state. I.e. the second order phase transition can't be realized. The only self-consistent variant is the first order phase transition. I.e., with fixed $m_Q/\la$ and increased $N_c$, there is {\it the point} within the region  $\mu_{\rm gl}(\mu\sim\la)\sim\la$ where the order parameter $\rho_{\rm higgs}$ drops from its value \eqref{(2.3)} {\it to zero}. And these reasonings are applicable to both types of phase transitions described in sections 3 and 4. The only difference is that either $\rho_{0}(N_c,\,m_Q/\la)$ changes with varying $N_c$ at fixed $m_Q/\la$, or it changes with varying $m_Q/\la$ at fixed $N_c$ and $N_F$.

\section{Conclusions}

The conclusion of this paper about the phase transition between the confinement and higgs regimes is opposite to the conclusion of the paper of E. Fradkin and S.H. Shenker [10] that the transition between these regimes is the crossover, not the phase transition. And although the theories considered in this paper
and in [10] are different ( see page 9 with the critique of the model used in \cite{FS}), an experience shows that there is a widely spread opinion that the conclusion of [10] is applicable to all QCD-like theories: both lattice and continuum, and both not supersymmetric and supersymmetric.  This opinion is in contradiction with the results of this paper.\\

Another types of phase transitions in  ${\cal N}=2$ SQCD are described in sections 6.1, 6.2, 7, 8 of \cite{ch17}.\\

I'm grateful to R.N. Lee for useful discussions about the gauge invariance.\\

\end{document}